\def\rfr#1{eq.(\ref{#1})}
\def\rfrs#1#2{eqs.(\ref{#1})-(\ref{#2})}
\def\Rfr#1{Eq.(\ref{#1})}

\def\dert#1#2{\frac{d#1}{d#2}}
\def\eqi{\begin{equation}}
\def\eqf{\end{equation}}
\def\eqia{\begin{eqnarray}}
\def\eqfa{\end{eqnarray}}
\def\rp#1#2{{#1\over#2}}
\def\ct#1{\cite{#1}}
\def\lb#1{\label{#1}}
\def\bm#1{{\mbox{\boldmath$#1$\unboldmath}}}

\def\cip{\cosh{\rp{g(t-t_0)}{c}}}
\def\sip{\sinh{\rp{g(t-t_0)}{c}}}
\def\tip{\tanh{\rp{g(t-t_0)}{c}}}
\def\cq{c^2}
\def\pare{\left(1+\rp{gx_0}{\cq}\right)}
\def\bet{\rp{v_0}{c}}
\def\gam{\sqrt{1-\left(\rp{V}{c}\right)^2}}
\def\spazio{-\rp{-\cq+c^4 C_1 C_2^2 e^{\rp{2g(t-t_0)}{c}}+2C_2 e^{\rp{g(t-t_0)}{c}}}
{g\left[-1+\cq C_1 C_2^2 e^{\rp{2g(t-t_0)}{c}}\right]}}
\def\velocità{\rp{2C_2 e^{\rp{g(t-t_0)}{c}}\left[1+c^2 C_1 C_2^2 e^{\rp{2g(t-t_0)}{c}}\right]}{c\left[-1+\cq C_1 C_2^2 e^{\rp{2g(t-t_0)}{c}}\right]^2}}
\def\primacost{-\rp{1+\left(\rp{gx_0}{\cq}\right)^2-\left(\rp{v_0}{c}\right)^2\left(1-\rp{2gx_0}{v_0^2\cq}\right)}{c^6\pare^4}}
\def\secondacost{\cq\rp{\pare^2}{\left(1+\rp{v_0}{c}+\rp{gx_0}{\cq}\right)}}
\def\pos{\rp{\cq}{g}\left[\rp{\pare^2}{\pare\cip-\rp{v_0}{c}\sip}-1\right]}
\def\vel{-c\rp{\pare^2\left[\pare\tip-\bet\right]}{\left[\pare-\bet\tip\right]^2}}
\def\temprop{\rp{c}{g}\rp{\sqrt{\left(1+\rp{gx_0}{\cq}\right)^2-\left(\rp{v_0}{c}\right)^2}
\pare\tip}{\left[\pare-\bet\tip\right]}}

\documentclass[11pt]{article}

\usepackage{amsmath,amsthm,amscd,amssymb}
\usepackage{latexsym}
\usepackage{graphics,graphicx}

\begin{document}

\noindent{\bf \LARGE{An analytical treatment of the Clock Paradox
in the framework of the Special and General Theories of Relativity
}}
\\
\\
\\
Lorenzo Iorio\\Dipartimento Interateneo di Fisica dell'
Universit${\rm \grave{a}}$ di Bari\\
INFN-Sezione di Bari
\\Via Amendola 173, 70126\\Bari, Italy
\\
\\

\begin{abstract}
In this paper we treat the so called clock paradox in an
analytical way by assuming that a constant and uniform force \bm F
of {\it finite} magnitude acts continuously on the moving clock
{\it along the direction of its motion} assumed to be {\it
rectilinear} (in space). No inertial motion steps are considered.
The rest clock is denoted as (1), the to--and--fro moving clock is
(2), the inertial frame in which (1) is at rest in its origin and
(2) is seen moving is $I$ and, finally, the accelerated frame in
which (2) is at rest in its origin and (1) moves forward and
backward is $A$. We deal with the following questions: I) {\it
What is the effect of the finite force acting on (2) on the proper
time intervals $\Delta\tau^{(1)}$ and $\Delta\tau^{(2)}$ measured
by the two clocks when they reunite? Does a differential aging
between the two clocks occur}, as it happens when inertial motion
and infinite values of the accelerating force is considered? The
Special Theory of Relativity is used in order to describe the
hyperbolic (in spacetime) motion of (2) in the frame $I$ II) {\it
Is this effect an absolute one}, i.e. does the accelerated
observer $A$ comoving with (2) obtain the same results as that
obtained by the observer in $I$, both qualitatively and
quantitatively, as it is expected? We use the General Theory of
Relativity in order to answer this question. It turns out that
$\Delta\tau_I=\Delta\tau_A$ for both the clocks,
$\Delta\tau^{(1)}$ and $\Delta\tau^{(2)}$ {\it do depend on}
$g=F/m$, and
$\Delta\tau^{(2)}/\Delta\tau^{(1)}=(\sqrt{1-\beta^2}{\rm
atanh}\beta)/\beta<1$. In it $\beta=V/c$ and $V$ is the velocity
acquired by (2) when the force inverts its action.
\end{abstract}

\section{Introduction}
In this paper we wish to quantitatively examine in detail the so
called clock paradox by accounting for the effects of the {\it
finiteness}  of the force $\bm F$ which accelerates and
decelerates the moving clock which will conventionally be denoted
as (2). The Special and General Theories of Relativity will be
used, as in \ct{gron} in which a symmetrical version of the clock
paradox has been considered. In it two clocks perform hyperbolic
motions with oppositely directed velocities and accelerations. The
role of finite acceleration in the twin paradox has recently been
investigated in \ct{nikolicfpl, vallis} without using the Einstein
theory of accelerated frames.

An unidimensional {\it rectilinear} (in space) path will be
considered in which the moving clock starts its motion with zero
initial velocity from the origin of the inertial frame $I$ in
which the rest clock, denoted conventionally as (1), is located.
After the velocity $V$ is reached, $\bm F$ is instantaneously
reversed.
Then, (2) is decelerated, stops and inverts its motion until the
velocity $-V$ is reached. At this point $\bm F$ is suddenly
reversed again, so that, when (2) stops, it meets (1) again and
they compare their readings.

In general, by assuming an infinite force (or no force at
all\footnote{Indeed, it is possible to imagine a situation with
one observer at rest and two oppositely moving inertial observers
which encounter each other at a certain spacetime event in which
no acceleration occurs at all. By accounting for the time `jump'
it is possible to obtain the desired absolute different aging at
the reunion with the rest clock \ct{olaf}. }) on the accelerated
clock, it can be obtained that (2) does lag behind the inertial
clock (1) when they reunite and compare their readings and that
this asymmetric effect is an {\it absolute} one. The paradoxical
symmetric outcome would come, instead, from an uncorrect
application of the Special Theory of Relativity to the motion of
(2) in the sense that the so called time `jump' in the time of (1)
as measured by (2) during its inertial motion must be considered
as well in order to obtain the desired result. Instead, if it is
not accounted for, the paradoxical symmetric situation comes out.

In this paper we wish to investigate what happens if, instead, a
{\it finite} force is considered: {\it does a differential aging
between the two clocks occur again? If so, what is the clock which
lags behind? What is the magnitude of this effect?} If the
relation between the proper times
  measured when they reunite again in order to compare the readings of their
  displays
  had to be considered as an absolute
  effect, both the inertial observer and the accelerated one should
  agree not only that a certain clock lags behind the other clock, but
  also the magnitude of such an effect should be the same. We will
  investigate quantitatively {\it if and how this feature is modified by accounting for the finite force experienced
  by (2) leading to some inconsistencies}. Indeed, the elapsed proper time is function only
of the observer's worldline and of its starting/ending points: it
is the Lorentzian length of the segment of worldline delimited by
given endpoints. The spacetime of an accelerated frame does not
present a real curvature as if a true gravitational field was
present. Whatever coordinates are used for flat spacetime, it will
always be flat. The coordinates used might change the form of the
metric, but they cannot create curvature. The choice of frame will
also change the coordinate expression of geodesics (e.g., the
worldline of clock (1)), but it will not change its geometrical
properties, including its proper-time length. Thus, the worldline
will curve in the sense that its spatial coordinates are not
constant in the accelerated frame, but it will still be straight
in the sense that it is inertial (and therefore an extremum of
proper-time length).

The motion will be described both in the inertial frame $I$, in
which (1) is at rest while (2) moves according to the special
relativistic hyperbolic (in spacetime) motion, and in the
accelerated frame $A$ comoving with (2) in which the latter one is
constantly at rest and (1) moves in a way which will be derived in
the framework of the General Theory of Relativity. In general, $t$
will denote the proper time of the clock which is at rest in a
given frame and $\tau$ the proper time of the clock which is seen
to be in motion in the same frame.

What is the experimental status of the clock paradox? Some
  experiments with elementary particles have been carried out until now.
  The celebrated experience of the
  circling muons at CERN \ct{muoncern} refers to a
  scenario in which the moving particles,
  following a $circular\ uniform$ motion, comes back periodically to the same
  point  where their lifetimes are measured
and compared to their proper lifetimes calculated at rest.
  Note that, in this case, {\it no
  acceleration along the direction of motion} exists as, instead, is the case of the
  rectilinear accelerated motion.
  In the case of a $rectilinear$ motion there
  is $no$ any experiment which tests a scenario like that involved
  in the to--and--fro journey of the {\it accelerated} clock; the
  observed dilation of the lifetime of muons in the cosmic rays
  which reach the sea level before decaying \ct{muonsealevel} does not imply a
  return of the muons to some point of reference.
  Moreover, no external force
  acts on the cosmic rays muons whose motion is rectilinear and
  uniform, i.e. an {\it one--way non---accelerated motion} occurs in
  this case which {\it cannot be assumed as an experimental test of the clock paradox in the case
  of rectilinear accelerated motion}.

\section{The point of view of the inertial clock}\lb{inclock}
\subsection{The special relativistic equation of motion of an accelerated particle in an inertial frame}
The equation of motion of a particle of mass $m$ acted upon by a
force $\bm F$, as viewed in an inertial frame $I$, is, according
to the Special Theory of Relativity \eqi\rp{d\bm p}{dt}=\bm
F,\lb{hyp}\eqf with \eqi \bm p=\rp{m\bm
v}{\sqrt{1-\left(\rp{v}{c}\right)^2}}.\nonumber\eqf Note that $t$
denotes the proper time of a standard clock located at the origin
of $I$.

Let us consider an unidimensional motion under the action of a
constant and uniform force $F$. A first integration of \rfr{hyp}
yields \eqi
\rp{v(t)}{\sqrt{1-\left[\rp{v(t)}{c}\right]^2}}=g(t-t_0)+C_1,\lb{yuhu}\eqf
with $g\equiv F/m$. \Rfr{yuhu} admits as solution \eqi x(t) =
\rp{\cq}{g}\sqrt{1+\left[\rp{g(t-t_0)}{c}+\rp{C_1}{c}\right]^2
}+C_2,\nonumber\eqf from which the velocity can be obtained
\eqi\rp{dx(t)}{dt} =
c\rp{\left[\rp{g(t-t_0)}{c}+\rp{C_1}{c}\right]}{\sqrt{1+\left[\rp{g(t-t_0)}{c}+\rp{C_1}{c}\right]^2}}.
\nonumber\eqf The constants of integration can be determined from
the initial conditions $x(t_0)=x_0$ and $v(t_0)=v_0$; then, \eqia
C_1 &=& \rp{v_0}{\sqrt{1-\left(\rp{v_0}{c}\right)^2}},\nonumber\\
C_2 &=& x_0-\rp{\cq}{g\sqrt{1-\left(\rp{v_0}{c}\right)^2}},
\nonumber\eqfa so that the complete characterization of the
special relativistic hyperbolic motion is \eqia
x-x_0&=&\rp{\cq}{g}\left\{\sqrt{1+\left[\rp{g(t-t_0)}{c}+\rp{v_0}{c\sqrt{1-\left(\rp{v_0}{c}\right)^2}}\right]^2
}-\rp{1}{\sqrt{1-\left(\rp{v_0}{c}\right)^2}}\right\},\lb{hyperspace}\\
\rp{d
x}{dt}&=&\rp{g(t-t_0)+\rp{v_0}{\sqrt{1-\left(\rp{v_0}{c}\right)^2}}}{\sqrt{1+\left[\rp{g(t-t_0)}{c}+\rp{v_0}{c\sqrt{1-\left(\rp{v_0}{c}\right)^2}}\right]^2
}},.\lb{hypervel}
\eqfa Note that, for $c\rightarrow\infty$,
\rfrs{hyperspace}{hypervel}
tend to \eqia x(t)&\rightarrow & x_0+v_0(t-t_0)+\rp{g(t-t_0)^2}{2},\nonumber\\
\rp{d x(t)}{dt}&\rightarrow & v_0+g(t-t_0).\nonumber\\
\eqfa
 The proper time interval of the moving particle can be
written as
\eqi\tau-\tau_0=\int_{t_0}^t\sqrt{1-\left[\rp{v(t^{'})}{c}\right]^2}dt^{'},\lb{gluglu}\eqf
according to the hypothesis of locality \ct{locality}.
It is interesting to note that \rfr{gluglu} does {\it not} depend
{\it explicitly} on the {\it acceleration}  of the moving
particle, but only on its velocity  which, however, contains the
force per unit mass $g$, as can be noted by \rfr{hypervel}.
\subsection{The motion of the clock (2) with respect to the clock (1)}
The results of the previous section will now be used in order to
describe the motion of the clock $(2)$ with respect to the clock
$(1)$. In this case $t$ is the proper time $\tau^{(1)}$ of the
clock $(1)$, which is at rest in $I$, while $\tau$ denotes the
proper time $\tau^{(2)}$ of the moving clock $(2)$. In the
following we will split the motion in four steps.
\subsubsection{From $t=t_0$ to $t=t_1$} 
In this stage
\begin{itemize}
  \item $g>0$
  \item $x(t)>0$
  \item $v(t)>0$
  \item $x(t_0)=0$
  \item $x(t_1)=\lambda_1$
  \item $v(t_0)=0$
  \item $v(t_1)=V$
\end{itemize}
i.e. the clock $(2)$ starts moving from the origin with zero
initial velocity and is accelerated to a velocity $V$ which is
reached at $t=t_1$ in $\lambda_1$. From \rfr{yuhu} it can be
obtained \eqi
t_1-t_0=\rp{V}{g\sqrt{1-\left(\rp{V}{c}\right)^2}}.\lb{unO}\eqf
Note that, for finite values of $g$, $t_1-t_0$ is finite as well;
the larger the force acting on (2), the shorter the time required
to reach $V$. From \rfr{hyperspace} and \rfr{unO} it can be
obtained the point at which the velocity $V$ is reached: it is
\eqi
\lambda_1=\rp{\cq}{g}\left[\rp{1}{\sqrt{1-\left(\rp{V}{c}\right)^2}}-1\right]\equiv\lambda.\lb{cvb}\eqf
Note that, for $g\rightarrow\infty$, it tends to zero; for finite
values of $g$ it is finite as well\footnote{The topic of lengths
in accelerated frames has recently been treated in \ct{nikolicajp,
tartrug}. }.

\Rfr{yuhu} in \rfr{gluglu} yields \eqi
\tau_1-\tau_0=\rp{c}{g}\ln\left\{\rp{g(t_1-t_0)}{c}+\sqrt{1+\left[\rp{g(t_1-t_0)}{c}\right]^2}\right\}\lb{ijk}.\eqf
Finally, \rfr{unO} in \rfr{ijk} allows to obtain \eqi
\tau_1-\tau_0=\rp{c}{g}{\rm
atanh}\left(\rp{V}{c}\right)
.\lb{UNo}\eqf Note that, {\it during the accelerated motion, the
proper time interval of the moving clock (2) is always shorter
than the proper time read by the inertial clock (1)}.
Note that, for $g\rightarrow\infty$,
both \rfr{unO} and \rfr{UNo} vanish.
%
%

%
%
\subsubsection{From $t=t_1$ to $t=t_2$}
In this stage
\begin{itemize}
  \item $g<0$
  \item $x(t)>0$
  \item $v(t)>0$
  \item $x(t_1)=\lambda$
  \item $x(t_2)=L$
  \item $v(t_1)=V$
  \item $v(t_2)=0$
\end{itemize}
i.e. at $t_1$ the same force as before is switched again on, but
in the opposite direction, so that the clock $(2)$ decelerates and
stops at $t_2$ in $x(t_2)=L$. From \rfr{yuhu} it can be obtained
\eqi t_2-t_1=\rp{V}{g\gam}=t_1-t_0.\lb{trE}\eqf \Rfr{trE} in
\rfr{hyperspace} yields \eqi L=\lambda+
\rp{\cq}{g}\left[\rp{1}{\sqrt{1-\left(\rp{V}{c}\right)^2}}-1\right].\lb{ELLE}\eqf
From \rfr{cvb} it follows \eqi
L=2\lambda=\rp{2\cq}{g}\left[\rp{1}{\sqrt{1-\left(\rp{V}{c}\right)^2}}-1\right].
\lb{ELLE}\eqf

The proper time interval of the clock (2) is, according to the
initial conditions of this stage \eqia\tau_2-\tau_1&=& \rp{c}{2g}
\ln\left\{\left[\rp{1+\left(\rp{V}{c}\right)}{1-\left(\rp{V}{c}\right)}\right]
\left[\rp{g(t_2-t_1)}{c}-\rp{V}{c\gam}+\right.\right.\nonumber\\
&+&\left.\left.\sqrt{\rp{1}{1-\left(\rp{V}{c}\right)^2}-\rp{2gV(t_2-t_1)}{\cq\gam}+\rp{g^2(t_2-t_1)^2}{\cq}}
\right]\right\}.\lb{megaformula}\eqfa
%
\Rfr{trE} in \rfr{megaformula} yields
\eqi\tau_2-\tau_1=\rp{c}{g}{\rm
atanh}\left(\rp{V}{c}\right)=\tau_1-\tau_0\lb{TRe}.
\eqf
\subsubsection{From $t=t_2$ to $t=t_3$} 
In this stage
\begin{itemize}
  \item $g<0$
  \item $x(t)>0$
  \item $v(t)<0$
  \item $x(t_2)=L$
  \item $x(t_3)=\lambda_2$
  \item $v(t_2)=0$
  \item $v(t_3)=-V$
\end{itemize}
i.e. the force continues to act upon the clock $(2)$ along the
negative $x$ axis so that it starts accelerating until velocity
$-V$ is reached at $t_3$. From \rfr{yuhu} it can be obtained \eqi
t_3-t_2=\rp{V}{g\sqrt{1-\left(\rp{V}{c}\right)^2}}=t_2-t_1=t_1-t_0.\lb{quattrO}\eqf
\Rfr{quattrO} in \rfr{hyperspace} yields for $\lambda_2$, i.e. the
place where $F$ is reversed \eqi
\lambda_2=-\rp{\cq}{g}\left[\rp{1}{\sqrt{1-\left(\rp{V}{c}\right)^2}}-1\right]+L=\lambda.\eqf
Note that, for $g\rightarrow\infty$, it tends to $L$.
 \Rfr{yuhu} in \rfr{gluglu} yields
\eqi
\tau_3-\tau_2=\rp{c}{g}\ln\left\{\rp{g(t_3-t_2)}{c}+\sqrt{1+\left[\rp{g(t_3-t_2)}{c}\right]^2}\right\}\lb{olp}.\eqf
Finally, \rfr{quattrO} in \rfr{olp} allows to obtain \eqi
\tau_3-\tau_2=\rp{c}{g}{\rm
atanh}\left(\rp{V}{c}\right)=\tau_2-\tau_1=\tau_1-\tau_0.\lb{QUATTRo}\eqf
\subsubsection{From $t_3$ to $t_4$}\lb{klop}
In this stage
\begin{itemize}
  \item $g>0$
  \item $x(t)>0$
  \item $v(t)<0$
  \item $x(t_3)=\lambda$
  \item $x(t_4)=0$
  \item $v(t_3)=-V$
  \item $v(t_4)=0$
\end{itemize}
i.e. the same force as before is switched again on, but in the
opposite direction, at $t_3$ and the clock $(2)$ is decelerated
until it stops at $t_4$ when it meets the clock $(1)$ again. From
\rfr{yuhu} it can be obtained \eqi
t_4-t_3=\rp{V}{g\gam}=t_3-t_2=t_2-t_1=t_1-t_0.\lb{seI}\eqf

The proper time interval of the clock (2) is \eqia\tau_4-\tau_3&=&
\rp{c}{2g}
\ln\left\{\left[\rp{1+\left(\rp{V}{c}\right)}{1-\left(\rp{V}{c}\right)}\right]
\left[\rp{g(t_4-t_3)}{c}-\rp{V}{c\gam}+\right.\right.\nonumber\\
&+&\left.\left.\sqrt{\rp{1}{1-\left(\rp{V}{c}\right)^2}-\rp{2gV(t_4-t_3)}{\cq\gam}+\rp{g^2(t_4-t_3)^2}{\cq}}
\right]\right\}.\lb{megaformula2}\eqfa \Rfr{seI} in
\rfr{megaformula2} yields \eqia\tau_4-\tau_3&=&\rp{c}{g}{\rm
atanh}\left(\rp{V}{c}\right)=\tau_3-\tau_2=\tau_1-\tau_0\lb{SEi}.\eqfa
\subsection{The total proper time intervals at the clocks' reunion}
The total proper time interval of the accelerated clock (2) at the
reunion with the rest clock (1), as viewed in $I$, is, then
\eqi\Delta\tau_I^{(2)}=\rp{4c}{g}{\rm
atanh}\left(\rp{V}{c}\right), \lb{zum}\eqf while the total proper
time interval of the rest clock (1) at the reunion with the moving
clock (2), as viewed in $I$, is
\eqi\Delta\tau_I^{(1)}=\rp{4V}{g\sqrt{1-\left(\rp{V}{c}\right)^2}}.\lb{ZOM}\eqf
Note that \rfrs{zum}{ZOM} {\it do depend on the magnitude of the
force per unit mass applied to (2). However, also in presence of
finite values of the force which acts upon the clock (2), it lags
always behind the rest clock (1)}.
\section{The point of view of the accelerated clock}
Does the observer comoving with (2) obtain the same results of (1)
in $I$ seen in the previous Section? It has been shown that it is
possible to preserve the absolute differential aging of the two
clocks when no accelerations at all occur (see, e.g. \ct{olaf}) by
using three inertial observers. What happens if a finite
acceleration felt by (2) is taken into account when the point of
view of (2) is considered? In this Section we will try to answer
this question by using the formalism of the General Theory of
Relativity.
\subsection{The equations of motion in an accelerated frame}
Let us consider the generic motion of a body with respect to an
inertial frame whose spacetime coordinates will be now denoted as
($X, Y, Z, T$). It is possible to construct a frame of reference
which is the relativistic analogue of a classical rigid\footnote{A
system of reference is called rigid if the space distance $\sigma$
between two reference points, as measured by standard
measuring--rods at rest in the system, is constant in time. In
this case $(d\sigma)^2=(dx)^2+(dy)^2+(dz)^2$ and the coordinates
$x, y, z$ are Cartesian and have a metric meaning.} frame of
Cartesian axes following the body in its motion, so that the
latter is constantly situated at the origin of this frame of
reference \ct{moller}. It will be denoted as $A$ and its spacetime
coordinates will be ($x, y, z, t$). It can be constructed from the
successive infinitesimal Lorentz transformations without rotations
of the spatial axes which determine the successive inertial frames
that are momentarily rest systems of the moving particle. It turns
out that the coordinate clock located at the origin is a standard
clock, i.e. $t=\tau^{'}$, where $\tau^{'}$ is the proper time of
the moving body on which the frame $A$ is constructed and which is
located at its origin. For a hyperbolic motion along the $x$ axis
the transformation between the inertial frame and $A$ is
  \eqia
  X&=&\rp{\cq}{g}\left(\cosh{\rp{gt}{c}}-1\right)+x\cosh{\rp{gt}{c}},\label{ics}\\
  Y&=&y,\ Z=z,\\
  T&=&\rp{c}{g}\sinh{\rp{gt}{c}}+\rp{x}{c}\sinh{\rp{gt}{c}}\label{ti}
  \eqfa
Note that, by using \eqia\cosh\alpha&\sim& 1+\rp{\alpha^2}{2},\nonumber\\
\nonumber \sinh\alpha&\sim& \alpha+\rp{\alpha^3}{6}\nonumber,
\eqfa it is possible to obtain, in the limit $c\rightarrow\infty$
\eqia
  X&\rightarrow& x+\rp{gt^2}{2},\nonumber\\
  T&\rightarrow&t\nonumber.
  \eqfa
Let us, now, derive the relation between the velocities of a
moving particle in $I$ and $A$. From \rfrs{ics}{ti} it is possible
to obtain \eqi
dX=dt\cosh\rp{gt}{c}\left[c\left(1+\rp{gx}{c^2}\right)\tanh\rp{gt}{c}+\rp{dx}{dt}\right],
\lb{deicx}\eqf \eqi dY=dy,\ dZ=dz,\eqf\eqi
dT=dt\cosh\rp{gt}{c}\left[\left(1+\rp{gx}{c^2}\right)+\rp{\tanh\rp{gt}{c}}{c}\rp{dx}{dt}\right]
\lb{deTi}.\eqf Then,
\eqi\rp{dX}{dT}=\rp{c\left(1+\rp{gx}{c^2}\right)\tanh\rp{gt}{c}+\rp{dx}{dt}}
{\left(1+\rp{gx}{c^2}\right)+\rp{\tanh\rp{gt}{c}}{c}\rp{dx}{dt}}.\lb{velVEL}\eqf
In the limit $c\rightarrow\infty$ \rfr{velVEL} yields the
Newtonian result for the motion in rectilinearly accelerated frame
\eqi\rp{dX}{dT}\rightarrow gt+\rp{dx}{dt}\equiv v_{\rm abs}=v_{\rm
trasc}+v_{\rm rel}.\eqf
 By using \rfrs{ics}{ti} in
\eqi(cdT)^2-(dX)^2-(dY)^2-(dZ)^2\nonumber\eqf it is possible to
obtain the spacetime interval of the accelerated frame $A$ which
is \eqi
(ds)^2=(cdt)^2\left(1+\rp{gx}{c^2}\right)^2-(dx)^2-(dy)^2-(dz)^2\lb{ds}.\eqf

Let us, now, consider the motion of a particle of mass $m$ in such
a spacetime. Its equation of motion with respect to the frame $A$
can be obtained, e.g., with the Lagrangian approach from
\eqi\rp{d\left(\rp{\partial\mathcal{L}}{\partial\dot x
}\right)}{d\tau}-\rp{\partial \mathcal{L}}{\partial
x}=0\nonumber\eqf and \eqi \mathcal{L}=\rp{m}{2}g_{\mu\nu}\dot
x_{\mu}\dot x_{\nu}.\nonumber\eqf The overdot means derivation
with respect to the proper time $\tau$ of the moving particle. For
an unidimensional motion along the $x$ axis it can, then, be
obtained
\eqi\rp{d^2x}{d\tau^2}=-g\left(1+\rp{gx}{c^2}\right)\left(\rp{dt}{d\tau}\right)^2\lb{eqz}\eqf
where $t$ is the time of the standard clock located at the origin
of the accelerated frame $A$. It is possible to express \rfr{eqz}
in terms of $t$ by noting that $\tau$ is given, in general, by
\eqi d\tau=\rp{\sqrt{g_{\mu\nu}dx^{\mu}dx^{\nu}}}{c}.\nonumber\eqf
\Rfr{ds} yields \eqi
d\tau=\sqrt{\left(1+\rp{gx}{c^2}\right)^2-\left(\rp{v}{c}\right)^2}dt,\lb{tauu}\eqf
where $v^2=(dx/dt)^2$. Note that the proper time of the moving
particle does not depend explicitly on the acceleration; it
depends on the position and on the velocity. Of course, they will
come from the solution of the equation of motion which account for
the effects of the force $F$. With \rfr{tauu} \rfr{eqz} becomes
\eqi\rp{d^2x}{dt^2}=\rp{2g}{\cq\left(1+\rp{gx}{\cq}\right)}\left(\rp{dx}{dt}\right)^2-g\left(1+\rp{gx}{\cq}\right)\lb{accel}.\eqf
It may interesting to note that for $c\rightarrow\infty$
\rfr{accel} reduces to the Newtonian equation
\eqi\rp{d^2x}{dt^2}=-g\nonumber.\eqf In order to solve
\rfr{accel}, let us pose \eqi p=\rp{dx}{dt};\lb{sosti}\eqf then
\eqi \rp{d^2 x}{dt^2}=p\dert{p}{x}\nonumber.\eqf \Rfr{accel} can,
then, be cast into the form
\eqi\dert{p}{x}=\rp{2g}{\cq\left(1+\rp{gx}{\cq}\right)}p-\rp{g}{p}\left(1+\rp{gx}{\cq}\right)\lb{aux}\eqf
whose general solution is \eqi p(x)=\pm
c\left(1+\rp{gx}{c^2}\right)\sqrt{1+C_1
c^6\left(1+\rp{gx}{c^2}\right)^2}.\lb{primostep}\eqf By
inserting\footnote{The $+$ sign must be retained with $dx>0$ and
vice versa because $dt>0$.} \rfr{primostep} in \rfr{sosti} allows
to obtain \eqi\rp{dx}{\left(1+\rp{gx}{\cq}\right)\sqrt{1+C_1
c^6\left(1+\rp{gx}{\cq}\right)^2}}=cdt,\nonumber\eqf whose general
solution is \eqi x(t)=\spazio.\lb{spazio}\nonumber\eqf The
velocity of the particle is \eqi
\rp{dx(t)}{dt}=\velocità.\lb{velocità}\nonumber\eqf The constants
of integration $C_1$ and $C_2$ can be determined from the initial
conditions $x(t_0)=x_0$, $dx(t_0)/dt=v_0$. It turns out that \eqia
C_1 & =& \primacost,\nonumber \\ C_2 &=& \secondacost. \nonumber
\eqfa Then, the position and the velocity of the particle are
given by \eqia
x(t)&=&\pos,\lb{sp}\\
\rp{dx(t)}{dt}&=&\vel \lb{vl}.
\eqfa Note that, for
$v_0=0$, \rfrs{sp}{vl} reduces to (8.173) and (8.174) of
\ct{moller}. Moreover, by expanding the hyperbolic functions to
order $\mathcal{O}(g)$, it can also be noted that, for
$g\rightarrow 0$,
\eqia x(t)&\rightarrow & x_0+v_0(t-t_0),\nonumber\\
\rp{dx(t)}{dt}&\rightarrow & v_0\nonumber.
\eqfa Moreover, for
$c\rightarrow\infty $, the Newtonian limit is restored
\eqia x(t)&\rightarrow & x_0+v_0(t-t_0)-\rp{g(t-t_0)^2}{2},\nonumber\\
\rp{dx(t)}{dt}&\rightarrow & v_0-g(t-t_0)\nonumber.
\eqfa
Concerning the proper time of the moving particle, \rfrs{sp}{vl}
in \rfr{tauu} yield \eqi
\tau-\tau_0=\temprop.\lb{tempoproprio}\eqf
\subsection{The motion of the clock (1) with respect to the clock (2)}
The results of the previous section will now be used in order to
describe the motion of the clock $(1)$ with respect to the clock
$(2)$. In this case $t$ is the proper time $\tau^{(2)}$ of the
clock $(2)$, which is now constantly at rest in the origin of $A$,
while $\tau$ denotes the proper time $\tau^{(1)}$ of the moving
clock $(1)$.

In regard to the proper time of (2), \rfrs{ics}{ti} yield \eqi
\tanh\rp{gt}{c}=\rp{gT}{c\left(1+\rp{gX}{c^2}\right)}.\lb{aiuto}\eqf
\Rfr{aiuto} allows to use the results of Section \ref{inclock}. By
using \rfr{unO} and \rfr{cvb} for all the four steps in which we
have subdivided the motion, it is straightforward to obtain
\eqi\Delta\tau_A^{(2)}=\rp{4c}{g}{\rm
atanh}\left(\rp{V}{c}\right)=\Delta\tau_I^{(2)}.\eqf This result
is very important because it shows that {\it the proper time
interval measured by the accelerated clock (2) after its reunion
with the clock (1) is the same both in $A$ and in $I$}, as it can
be expected.

The calculation of the proper time of the clock (1) is a bit more
involved. To this aim, let us note that in \rfr{velVEL} it is
always $dX/dT=0$ for the clock (1), of course; then \eqi
\rp{dx}{dt}=-c\left(1+\rp{gx}{\cq}\right)\tanh\rp{gt}{c}\lb{vvl}\eqf
must always hold. We will split the motion in four steps.
\subsubsection{From $t=t_0$ to $t=t_1$}\lb{startsec} 
In this stage the relevant initial conditions are
\begin{itemize}
  \item $g>0$
  \item $x(t_0)=0$
  \item $v(t_0)=0$ (from \rfr{velVEL} for $t=0$ and $dX/dT= 0$)
\end{itemize}
From \rfr{tempoproprio} one gets \eqi
\tau_1-\tau_0=\rp{c}{g}\tanh{\rp{g(t_1-t_0)}{c}}\lb{tippo}. \eqf
By using \rfr{UNo} for $t_1-t_0$ in \rfr{tippo} it is possible to
obtain
\eqi\left.\Delta\tau_A^{(1)}\right|_{0-1}=\rp{V}{g}\lb{PRIMA}.\eqf

Note that, from \rfr{vvl} for $g>0$ and $t=t_1$ of \rfr{UNo}, it
turns out \eqi
v(t_1)=-V\left(1+\rp{gl_1}{\cq}\right).\lb{incond}\eqf

\subsubsection{From $t=t_1$ to $t=t_2$}\lb{turnsec1}
In this stage the relevant initial conditions are
\begin{itemize}
  \item $g<0$
  \item $x(t_1)=-l_1$
  \item $v(t_1)=-V\left(1+\rp{gl_1}{c^2}\right)$
\end{itemize}
i.e. at $t_1$ the same inertial force as before is switched again
on but in the opposite direction, so that the clock $(1)$
decelerates until it stops.

The proper time interval of the clock $(1)$ can be obtained from
\rfr{tempoproprio} for $x_0=-l_1,\ v_0=-V(1+gl_1/c^2),\ g=-g$. It
is
\eqi\tau_2-\tau_1=\rp{c}{g}\rp{\left(1+\rp{gl_1}{\cq}\right)\sqrt{1-\left(\rp{V}{c}\right)^2}\tanh{\rp{g(t_2-t_1)}{c}}}
{1-\rp{V}{c}\tanh{\rp{g(t_2-t_1)}{c}}}.\lb{VAFFA}\eqf \Rfr{TRe}
for $t_2-t_1$ in \rfr{VAFFA} yields \eqi
\left.\Delta\tau_A^{(1)}\right|_{1-2}=\rp{V\left(1+\rp{gl_1}{\cq}\right)}{g\sqrt{1-\left(\rp{V}{c}\right)^2}}\lb{SECONDA}.\eqf
It is necessary to derive an explicit expression for $l_1$, as it
will become clear later. From \rfr{sp}, evaluated for $x=-l_1$,
$x_0=v_0=0$, and \rfr{UNo} for $t_1-t_0$ it follows \eqi
l_1=\rp{\cq}{g}\left[1-\sqrt{1-\left(\rp{V}{c}\right)^2}\right].\lb{elleuno}\eqf

Finally, note that from \rfr{vl}, evaluated for $t-t_0=t_2-t_1$ of
\rfr{TRe}, $g<0$ and $v_0$ of \rfr{incond}, it turns out that
$v(t_2)=0$.

\subsubsection{From $t_2$ to $t_3$}\lb{turnsec2}
In this stage the relevant initial conditions are\footnote{Note
that, since (1) and (2) are at rest relative to each other at the
inversion point, the maximum distance between the two clocks poses
no ambiguity and is the same when measured in $I$ and $A$; this is
the reason why the same symbol $L$ as before in Section
\ref{inclock} is used here. Moreover, note that from \rfr{ics} for
t=0 it follows just $x=X$. }
\begin{itemize}
  \item $g<0$
  \item $x(t_2)=-L$
  \item $v(t_2)=0$
\end{itemize}
i.e. the inertial force continues to act upon the clock $(1)$
along the positive $x$ axis so that it starts accelerating until a
velocity related to $V$ is reached at $t_3$.  The proper time
interval of $(1)$ can be obtained from \rfr{tempoproprio} for
$x_0=-L, v_0=0,
 g=-g$. It is \eqi
\tau_3-\tau_2=\rp{c}{g}\left(1+\rp{gL}{c^2}\right)\tanh{\rp{g(t_3-t_2)}{c}}.\lb{tpro}\eqf
\Rfr{QUATTRo} for $t_3-t_2$  and \rfr{ELLE} for $L$ in \rfr{tpro}
yield \eqi
\left.\Delta\tau_A^{(1)}\right|_{2-3}=\left(1+\rp{gL}{\cq}\right)\rp{V}{g}=\rp{2V}{g\sqrt{1-\left(\rp{V}{c}\right)^2}}
-\rp{V}{g}.\lb{TERZA}\eqf

Note that, from \rfr{vvl} for $g<0$ and $t=t_3$ of \rfr{QUATTRo},
it turns out \eqi v(t_3)=V\left(1-\rp{gl_2}{\cq}\right).\eqf

\subsubsection{From $t_3$ to $t_4$}\lb{endsec}
In this stage the relevant initial conditions are
\begin{itemize}
  \item $g>0$
  \item $x(t_3)=-l_2$
  \item $v(t_3)=V\left(1-\rp{l_2g}{\cq}\right)$
\end{itemize}
i.e. the same force as before is switched again on, but in the
opposite direction, at $t_3$ and the clock $(1)$ is decelerated
until it stops at $t_4$ when it meets the clock $(2)$ again.
 The proper time interval of the clock $(1)$ can
be obtained from \rfr{tempoproprio} for $x_0=-l_2,\
v_0=V(1-gl_2/\cq),\ g=g$. It is
\eqi\tau_4-\tau_3=\rp{c}{g}\rp{\left(1-\rp{gl_2}{\cq}\right)\sqrt{1-\left(\rp{V}{c}\right)^2}\tanh{\rp{g(t_4-t_3)}{c}}}
{1-\rp{V}{c}\tanh{\rp{g(t_4-t_3)}{c}}}\lb{tpr}\eqf \Rfr{SEi} for
$t_4-t_3$ in \rfr{tpr} yields
\eqi\left.\Delta\tau_A^{(1)}\right|_{3-4}=\rp{V\left(1-\rp{gl_2}{\cq}\right)}{g\sqrt{1-\left(\rp{V}{c}\right)^2}}.\lb{QUARTA}\eqf
The explicit expression for $l_2$ can be obtained from \rfr{sp},
evaluated for $x=0$, $x_0=-l_2,\ v_0=V(1-gl_2/\cq)$, and \rfr{SEi}
for $t_4-t_3$. Then, \eqi
l_2=\rp{\cq}{g}\left[1-\sqrt{1-\left(\rp{V}{c}\right)^2}\right]=l_1.
\lb{elledue}\eqf
\subsection{The total proper time interval of the clock (1)}
From \rfr{PRIMA}, \rfr{SECONDA}, \rfr{TERZA} and \rfr{QUARTA}, and
because $l_1=l_2$ according to \rfr{elleuno} and \rfr{elledue}, it
turns out
\eqi\Delta\tau_A^{(1)}=\rp{4V}{g\sqrt{1-\left(\rp{V}{c}\right)^2}}=\Delta\tau_I^{(1)}.\eqf
This is an important result because it shows that {\it the proper
time interval reckoned by the clock (1) after its reunion  with
the clock (2) is the same both in $I$ and in $A$}, as expected.
\section{Concluding remarks}\lb{summary}
 In this paper we have studied the clock paradox in the framework
 of the Special and General Theories of Relativity. We have
 considered a {\it rectilinear } (in space) motion of the moving clock during
 which it is continuously acted upon by a force which, at a certain instant, is
 reversed, slows it down until inverts its motion, re-accelerates it, inverts once more its action
 and decelerates again the moving clock
 until the latter one stops and meets again the rest clock.

  In the case of an uniform and constant
  force {\it of finite magnitude in the direction of the motion,
  such force does affect both the proper time of the moving clock and the proper time of the rest clock which sees it moving to--and--fro}.

 The expressions for the proper time intervals $\Delta\tau$ measured by a given clock at the clocks'
 reunion are the {\it same}, as expected, both in the inertial frame $I$ in which (1) is at rest while (2) performs a
 special
 relativistic hyperbolic (in spacetime) motion, and in the accelerated frame $A$ in which (2) is at
 rest and
 the General Theory of Relativity has been adopted in order to describe the motion of
 (1), i.e. $\Delta\tau^{(1)}_I=\Delta\tau^{(1)}_A$ and
 $\Delta\tau^{(2)}_I=\Delta\tau^{(2)}_A$.

 It turns out that
 \eqi\rp{\Delta\tau^{(2)}}{\Delta\tau^{(1)}}=\rp{\sqrt{1-\beta^2}{\rm atanh}\beta}{\beta}<1\nonumber,\eqf
 where $\beta\equiv V/c$,
 i.e., {\it the moving clock lags always behind the rest clock by an
 amount which depends only on the speed reached by the moving clock when the force inverts
 its action}. This result agrees with that found by the authors of
 \ct{gron}: indeed,
 according to them, the clock with the grater acceleration will
 mark shorter proper time intervals than the clock with smaller
 acceleration when they meet again.

In conclusion, in the case of a purely accelerated motion of the
clock which moves to--and--fro along a spatial straight line, a
differential aging with respect to the rest clock takes place. The
moving clock lags always behind the rest clock by an amount which
is different with respect to that which occurs when only inertial
motion is considered. The Special and General Theories of
Relativity are able to explain in a consistent way this feature.
\section*{Acknowledgments}
I am profoundly  grateful to F. Selleri  who motivated and
encouraged this study with insightful discussions and
observations. Thanks also to H. Nikoli\'{c} for interesting
remarks and references and to ${\rm \O}$. Gr${\rm \o}$n for the
reference of his paper.

\end{document}